\newcommand{\arxivVersion}{\True}
\documentclass[runningheads,letterpaper]{include/llncs}

\usepackage{graphicx}
\usepackage[tight,footnotesize]{subfigure}
\usepackage{tikz}
\usetikzlibrary{arrows,chains,matrix,positioning,scopes}
\usepackage{amsmath}
\usepackage{amssymb}
\usepackage[b]{esvect}  % vector
\usepackage{enumerate}
\usepackage{float}

\usepackage{pdfpages}
\usepackage{pgfplots}
\usetikzlibrary{plotmarks}
\usepackage{multirow}
\usepackage[nodayofweek]{datetime}
\usepackage{stmaryrd}
\usepackage{wasysym}
\usepackage{verbatim}   % for \begin{comment}
\newcommand{\hilightcolor}{gray} % used in the text!
\newcommand{\hilight} [1]{\setlength{\fboxsep}{1.0pt}\colorbox{\hilightcolor!30}{#1}} % blue
\newcommand{\hilightg}[1]{\setlength{\fboxsep}{1.0pt}\colorbox{\hilightcolor!30}{#1}}
\newcommand{\hilightr}[1]{\setlength{\fboxsep}{1.0pt}\colorbox{\hilightcolor!30}{#1}} % red

\usepackage{url}

\usepackage[noend]{algorithmic}
\usepackage{algorithm}
\usepackage{syntax, etoolbox}
\newcommand{\CASE}[1]{\STATE \textbf{case} #1\textbf{:} \begin{ALC@g}}
\newcommand{\ENDCASE}{\end{ALC@g}}

\newcommand{\DEFAULT}{\STATE \textbf{default:} \begin{ALC@g}}
\newcommand{\ENDDEFAULT}{\end{ALC@g}}
\newcommand{\DEFAULTLINE}[1]{\STATE \textbf{default:} }

\usepackage{pgfplots}
\usetikzlibrary{plotmarks}
\usepackage{stmaryrd}
\usepackage{wasysym}
\usepackage{tabularx}
\usepackage{listings}
\newcommand{\localCommand}{} % command names start with \
\newcommand{\newlengthsettowidth}[2]{\newlength {#1} \settowidth{#1}{#2}}

\newcommand{\ensurecommand}[2]{
  \providecommand{#1}{}
  \renewcommand{#1}{#2}}

\newcommand{\atl}{\geq}                                 % at least
\newcommand{\atm}{\leq}                                 % at most
\newcommand{\union}       {\mathbin        {\cup}}      % set union
\newcommand{\Union}       {\operatorname{\bigcup}}      % big union
\ensurecommand{\subsetneq}  {\subsetneqq}               % strict subset, violating arithmetic tradition, but unambiguous. Avoid if possible
\newcommand{\Ifthenelse}[3]{\ifthenelse{#1}{#2}{#3}}   % for consistency: initial capital
\newcommand{\Ifthen}    [2]{\Ifthenelse{#1}{#2}{}}
\newcommand{\Unless}    [2]{\Ifthen{\not {#1}}{#2}}
\newcommand{\Equal}     [2]{\equal{#1}{#2}}            % for consistency: initial capital
\newcommand{\Empty}     [1]{\Equal{#1}{}}
\newcommand{\True}         {\Equal{1}{1}}
\newcommand{\False}        {\Equal{1}{2}}

\newcommand{\renewcounter}[4]{\renewcommand{#1}{(#4 {#3})}\renewcommand{#2}{(#4 {#3})}}

\newcommand{\alphaenumi}{\renewcounter{\theenumi}{\labelenumi}{enumi}{\alph}}  % (a) ... (e)
\newdateformat{abbrevdate}{\shortmonthname\ \twodigit{\THEDAY}, \THEYEAR}

\providecommand{\Draftmode}{\False}

\newcommand{\Itedraft}    [2]{\Ifthenelse{\Draftmode}{#1}{#2}}
\newcommand{\Ifdraft}     [1]{\Itedraft{#1}{}}

\newcommand{\drafttext}   [2][\draftcolor]{\Ifdraft{{\color{#1}#2}}}
\newcommand{\draftpointer}{\makebox[0mm][c]{$^*$}}
\newcommand{\draftmargin} [2][\draftcolor]{\drafttext[#1]{\draftpointer\marginpar[\raggedleft\small{\color{#1}#2}]{\raggedright\small{\color{#1}#2}}}} % marginalia, shown in draftmode only
\newcommand{\draftdate}   [1][\abbrevdate\today, \currenttime]{\drafttext{\makebox[0mm][l]{\normalfont\tiny \ \ (Draft of #1)}}} % use behind title or author names
\newcommand{\draftauthor} [3]{
  \newcommand{#1}[1]{\drafttext  [#3]{##1}}   % Use:     \draftauthor{\thomas}{\thomasmargin}{green}
  \newcommand{#2}[1]{\draftmargin[#3]{##1}}}  % defines: \thomas{long comment}, \thomasmargin{short comment}

\newcommand{\useifspacecommand}{\draftauthor{\ifspace}{\ifspacemargin}{lightgray}}

\Ifdraft{\setlength{\overfullrule}{10mm}}

\newcommand{\C}{\texttt C}

\newcommand{\centerTwoOut}  [2]               {#1 \hfill #2}                          % |#1      #2|. Use with  two  large objects.
\newlength{\posLength}

\newlengthsettowidth{\tabLength}{\ \ \ }
\newcommand{\wbox}[2][\tabLength]{\hspace*{#1}\mbox{#2}\hspace*{#1}}

\newcommand{\0}[1]{}                                                   % comment from { to }. Note: {} is a token, so use A\0{ } B for ONE blank: A B
\newcommand{\End}{\end{document}}                                      % ignore rest of document
\newcommand{\format}{\drafttext{\underline{\makebox[0mm][c]{$\mid$}}}} % Mark commands used for final formatting: \format\vspace{2mm}. Also creates visible mark in text (of zero width)
\newcommand{\emphasize}[1]{\textbf{#1}}                                % stronger than \emph
\newcommand{\refcapital}[3]{#1#3} % capitalized: Definition/Def.

\newcommand{\reffull}[2]{#1} % full name: Definition/definition
\renewcommand{\reffull}[2]{#2} % abbreviated name: Def./def.

\newcommand{\appendixref}   [2][!]{\genericref[#1] A a {\reffull{ppendix}   {pp.}} {appendix}   {#2}}

\newcommand{\figureref}     [2][!]{\genericref[#1] F f {\reffull{igure}     {ig.}} {figure}     {#2}}

\newcommand{\lineref}       [2][!]{\genericref[#1] L l {\reffull{ine}       {ine}} {line}       {#2}}

\newcommand{\sectionref}    [2][!]{\genericref[#1] S s {\reffull{ection}    {ect.}}{section}    {#2}}
\newcommand{\tableref}      [2][!]{\genericref[#1] T t {\reffull{able}      {able}}{table}      {#2}}

\newcommand{\schemeref}     [2][!]{\genericref[#1] S s {cheme} {scheme}     {#2}} % I spelled this out, since ``Sch.'' is hard to understand

\newcommand{\equationref}[2][!]{\Ifthen{\Equal{#1}!}{\refcapital E e {\reffull{quation} {q.}}~}(\ref{equation: #2})}

\newcommand{\genericref}   [6][!]{\Ifthen{\Equal{#1}{!}}{\refcapital{#2}{#3}{#4}~}\ref{#5: #6}} % args: [!]{A}{a}{ppendix}{appendix}{Proofs}
\newcommand{\PProof}{\textbf{Proof}}                     % introducing a proof

\algsetup{indent=1.5em}

\newcommand{\plkeyword}[1]{\textbf{#1}}

\newcommand  {\FOREACH}{\FORALL}

\renewcommand{\algorithmicforall} {\plkeyword{for each}}
\newcommand{\code}[1]{\texttt{#1}}

\newcommand{\shortrange}[2]{\{#1..#2\}} % {1..n} (save some space compared to {1,...,n}

\draftauthor{\thomas} {\thomasmargin}{cyan}
\draftauthor{\peizun} {\peizunmargin}{blue}
\useifspacecommand

\AtBeginEnvironment{grammar}{\small}
\newcommand{\bnf}[1]{\emph{#1}}
\newcommand{\terminalfont}[1]{\textbf{#1}}
\newcommand{\terminal}    [1]{\terminalfont{#1}}
\newcommand{\nonterminal} [1]{\bnf{#1}}
\newcommand{\locals}{L}
\newcommand{\shares}{G} % changed to G (``global'') since our programs are not necessarily threaded
\newcommand{\numlocals}{|\locals|}
\newcommand{\numshares}{|\shares|}

\newcommand{\vars}{V}
\newcommand{\localvars}{\vars_{\locals}}
\newcommand{\sharedvars}{\vars_{\shares}}
\newcommand{\numlocalvars}{|\localvars|}
\newcommand{\numsharedvars}{|\sharedvars|}

\newcommand{\threadstate}[2]{(#1,#2)}

\newcommand{\Minf}{M^\infty}
\newcommand{\Sinf}{S^\infty}
\newcommand{\transrelinf}{\rightarrow}

\newcommand{\nondet}{\star}

\newcommand{\StepMath}[1][]{\textsf{image}\Unless{\Empty{#1}}{_{#1}}} % careful'': changed to 'image'

\newcommand{\B}{\mathcal{B}}

\newcommand{\pc}{\mathit{pc}}

\newcommand{\stmt}{\mathit{stmt}}

\newcommand{        \covers}   {\succeq}

\newcommand{\query} Q

\usepackage{pifont}% http://ctan.org/pkg/pifont
\newcommand{\cmark}{\text{\ding{51}}}%
\newcommand{\xmark}{\text{\ding{55}}}%
\newcommand{\LOC}{\emph{LOC}}
\newcommand{\YES}{\cmark}%{\CIRCLE}
\newcommand{\NO} {\xmark}%{\Circle}

\newcommand{\ourapi}     {\textsc{Ijit}}
\newcommand{\ourapiplain}     {IJIT}
\newcommand{\apiotf}  {{\sc  jit}}
\newcommand{\apibp}   {{\sc bp}}
\newcommand{\tts}   {{\sc tts}}

\newcommand{\ECUT}  {\textsc{Ecut}}

\newcommand{\KM}    {\textsc{Km}}
\newcommand{\AKM}    {\textsc{Akm}}

\newcommand{\SATABS}{\textsc{Sat\-Abs}}
\newcommand{\SLAM}  {\textsc{Slam}}

\newcommand{\BWS}   {\textsc{Bws}}

\newcommand{\spin}   {\textsc{Spin}}

\newcommand{\sstops}{f} % system state to program state
\newcommand{\pstoss}{f^{-1}}

\newcommand{\version}[2]{#1(#2)}

\begin{document}
\mainmatter  % start of an individual contribution
%===================================Title======================================%

\newcommand{\theTitle}{%
  %Unbounded-Thread
  % Multi-Threaded 
  \ourapiplain: An API for Boolean Program Analysis \\
  with Just-in-Time Translation\Ifthen{\arxivVersion}{\\ (Extended Technical Report)}}

\newcommand{\theTitlePlain}{\ourapiplain: An API for Boolean Program Analysis with Just-in-Time Translation}

\title{\theTitle\thanks{This work is supported by NSF grant no.~1253331.}\draftdate}

\titlerunning{\theTitlePlain}

%=================================Authors=======================================%
\author{Peizun Liu
  %\inst 1 % only one inst., hence no (ugly) superscripts needed
  \and Thomas Wahl
  %\inst 1
}

\authorrunning{\theTitlePlain} % feature abused for this document to repeat the title also on left hand pages

%=====================================Institutes===============================%
\institute{Northeastern University, Boston, United States}

%==================================Make Title==================================%
\toctitle{Peizun Liu's LNCS Template}
\tocauthor{Peizun Liu based on LNCS}
\maketitle

%================================= Abstract====================================%
\begin{abstract}
  Exploration algorithms for explicit-state transition systems are a core
  back-end technology in program verification. They can be
  applied to \emph{programs} by generating the transition system on the fly,
  avoiding an expensive up-front translation. An on-the-fly strategy
  requires significant modifications to the
  implementation, into a form that stores states directly as valuations of
  program variables. Performed manually on a per-algorithm basis, such
  modifications are laborious and error-prone.

  In this paper we present the \ourapi\ Application Programming Interface
  (API), which allows users to automatically transform a given transition
  system exploration algorithm to one that operates on \emph{Boolean} programs.
  The API converts system states temporarily to
  program states \emph{just in time} for expansion via image
  computations, forward or backward. 
  Using our API, we have effortlessly extended various non-trivial
  (e.g.\ infinite-state) model checking algorithms to operate on
  multi-threaded Boolean programs. We demonstrate the ease of use of the
  API, and present a case study on the impact of the just-in-time
  translation on these algorithms.
\end{abstract}

%\keywords{multi-threaded Boolean programs, on-the-fly}

%====================================Main Body=================================%

\section{Introduction}
\label{section: Introduction}

\emph{Boolean programs} \cite{BR00a}, a finite-data abstraction of
general-purpose software obtained by predicate abstraction \cite{GS97},
have proved to be an intermediate notation very useful for verification
that factors out the data complexity from programs\Ifthen{\arxivVersion}{, such as
  (unbounded) integers or dynamic data structures, while leaving the
  control structure intact}.\Ifthen{\arxivVersion}{\ Abstraction refinement techniques
  have been developed to adjust the precision with which data flow in
  programs is retained to a level just sufficient to prove properties of
  interest, or reveal genuine errors \cite{BMMR01,CKSY04,JM06}.} State
exploration algorithms, however, are typically designed to operate on forms
of transition systems. To apply these algorithms to Boolean programs, one
can in principle translate the input program into a transition system,
before starting the exploration. This input translation incurs, however, a
blow-up that is exponential in the number of program variables.

This classic problem in program verification has led to sophisticated
algorithms that translate the program into a transition system \emph{on the
  fly}, as the state space is explored. This idea was pioneered for model
checking algorithms by the \spin\ tool~\cite{GH97}. In general, to convert
an exploration algorithm into an on-the-fly version, the state
representation data structure needs to be changed everywhere in the
implementation to a tuple over program variable valuations. Consequently,
operations on the state representation, notably image computations, need to
be re-implemented as well, to reflect the program semantics.

Such an algorithm re-implementation avoids the exponential
pro\-gram-to-tran\-si\-tion-system translation, but comes with its own
cost: due to its low-level nature, it is laborious and error-prone,
especially for sophisticated algorithms. In the rest of this paper we
describe a way to \emph{automatically} construct on-the-fly program state
explorers from implementations operating on transition systems. We leave
the system state data structure intact (hence no algorithm
re-implementation), and pass the Boolean program as input (hence no input
program translation).
Our strategy is then as follows: whenever predecessor or successor images
need to be computed, the current system state is converted temporarily and
\emph{just in time} for the image computation into a Boolean program state.
The image is then computed using the program execution semantics, e.g.\ via
pre- or post-conditions. The resulting image states are converted back to,
and stored as, system states. This process is repeated for each image
computation.

This simple strategy has one crucial advantage: it requires very little
change on a per-algorithm basis: once we have provided image operations for
Boolean programs (a one-time effort), all we need to do is replace the
calls to image functions in the original implementation by new functions
that take a system state and (i) convert it to a Boolean program state,
(ii) apply the image, and (iii) convert the result back. These steps can be
encapsulated into a single operation.

Being largely independent of the underlying algorithm, this strategy can be
automated. To this end, we present an Application Programming Interface
(API) that provides conversion functions between system and Boolean program
states. It further offers implementations of common image operations on
Boolean programs, including standard pre- and post-images, as well as more
complex image operations for infinite-state system exploration. Our API
permits users to transform a wide range of transition system exploration
algorithms into Boolean program versions automatically---with little effort
and a high degree of reliability---, including sophisticated reachability
and coverability algorithms for infinite-state systems such as Petri nets.

For an experimental case study, we have implemented several exploration
algorithms in three versions: (a) one that uses the naive \emphasize{input
  translate} option, (b) one that implements the manual
\emphasize{algorithm re-implement} option, and (c)~one that uses our API to
perform \emphasize{just-in-time translation}. The comparison (c) against
(b) demonstrates that the repeated state representation conversion is not
harmful: using our API we achieve almost the same efficiency as the gold
standard of re-implementation by hand. The comparison (c) against (a)
demonstrates that the just-in-time version is vastly more efficient than
the version employing up-front input translation.

\section{Boolean Programs and Thread-Transition Systems}
\label{section: Boolean Programs and Thread-Transition Systems}

Our API allows exploration algorithms that operate on transition systems
derived from \emph{Boolean programs} (BP) \cite{BR00a} to be applied
directly to such programs, circumventing the blow-up incurred by the input
translation. In this section we formalize the language of (possibly
threaded) BPs and the transition system model of \emph{thread transition
  systems} \cite{KKW14a}. The latter serve as the input language of
exploration algorithms that we later wish to apply directly to BPs.

\subsection{Boolean Programs}
\label{section: Boolean Programs}

Boolean programs typically arise from predicate abstractions of application
code in system-level languages. All variables are of type \code{bool}.
Control flow constructs are optimized for synthesizability and therefore
include ``spaghetti statements'' like \terminal{skip} and \terminal{goto}.
An overview of the syntax of BPs is given in \figureref{boolean program
  syntax}. A~program consists of a \terminal{decl}aration of \emph{global}
Boolean variables, followed by a list of functions. A function consists of
a \terminal{decl}aration of \emph{local} Boolean variables, followed by a
list of labeled statements.
% \vspace{-13pt}
\begin{figure}[htbp]
  \setlength{\grammarparsep}{5pt plus 1pt minus 1pt} % increase separation between rules
  \setlength{\grammarindent}{5.5em} % increase separation between LHS/RHS
  \centering
%  \fbox
  {%
    \begin{minipage}{0.40\textwidth}
      \vspace{0.1pt}
      \begin{grammar}{\small}
        <prog> ::= \terminal{decl} \nonterminal{varlist}\terminal; \nonterminal{func}$^*$
      \end{grammar}

      \begin{grammar}{\small}
        <stmt> ::= \nonterminal{seqstmt}
          \alt \terminal{start_thread} \nonterminal{label}
          \alt \terminal{end_thread}
          \alt \terminal{atomic} \terminal{\{} [\nonterminal{stmt}\terminal;]$^*$ \terminal{\}}
          \alt \terminal{wait}
          \alt \terminal{signal}
          \alt \terminal{broadcast}
      \end{grammar}
    \end{minipage}
   \begin{minipage}{0.59\textwidth}
      \begin{grammar}{\small}
        <func> ::= \terminal{void} \nonterminal{name} \terminal(\nonterminal{varlist}\terminal) 
          \terminal{begin} \\
          \format\hspace*{3mm} \terminal{decl} \nonterminal{varlist}\terminal; \\
          \format\hspace*{3mm} [\nonterminal{label}\terminal: \nonterminal{stmt}\terminal;]$^*$ \\
          \terminal{end}

        <seqstmt> ::= \terminal{skip}
          \alt \terminal{goto} \nonterminal{labellist}
          \alt \terminal{assume} \terminal(\nonterminal{expr}\terminal)
          \alt \nonterminal{varlist} \terminal{:=} \nonterminal{exprlist} [\terminal{constrain} \nonterminal{expr}]
          \alt \terminal{assert} \terminal(\nonterminal{expr}\terminal)
      \end{grammar}
    \end{minipage}
  }
  \caption{Boolean program syntax (partial)}
  \label{figure: boolean program syntax}
\end{figure}
% \vspace{-13pt}

We illustrate the intuition behind individual statements of BPs. Among the
sequential statements (\nonterminal{seq\-stmt}), \terminal{skip} advances
the program counter~(pc); \terminal{goto} \nonterminal{labellist}
nondeterministically chooses one of the given labels as the next~pc;
\format\linebreak \terminal{assume} terminates executions that do not
satisfy the given expression. Statement \terminal{:=} assigns, in parallel,
each value in the given \nonterminal{exprlist} to the respective variable
in the same-length \nonterminal{varlist}, but terminates the execution if
the result does not satisfy the \terminal{constrain} expression, if any.
Statement \terminal{assert} indicates assertions for verification and
otherwise acts like \terminal{skip}.
The meaning of function calls (possibly recursive) and return statements
is standard and omitted. In all cases, \nonterminal{expr} is a Boolean expression over
global and local program variables, the constants $0$ and~$1$, and the
choice symbol~$\nondet$~; the latter nondeterministically evaluates to $0$
or $1$.

In the presence of multiple threads, the global variables are \emph{shared}
(both read and write) between the threads. The executing thread is called
\emph{active}, the others \emph{passive}. All sequential statements have
asynchronous semantics, i.e.\ they change the local variables of only the
active thread. The other statements in \figureref{boolean program syntax}
intuitively behave as follows:
\begin{description}
\item[\normalfont\terminal{start_thread} \nonterminal{label}] (i) advances
  the program counter of the executing thread, and (ii) creates a new
  thread whose local variables are copied from the executing thread and
  whose pc is given by \nonterminal{label};

\item[\normalfont\terminal{end_thread}] terminates the executing thread;

\item[\normalfont\terminal{atomic} $\{$\nonterminal{stmt}$^*\}$] denotes
  atomic execution: a thread executing inside an atomic section cannot be
  preempted;

\item[\normalfont\terminal{wait}] blocks the execution of a thread (see
  next);

\item[\normalfont\terminal{signal}] advances the pc of the executing thread
  and nondeterministically wakes up \emph{one} thread
  blocked at a \terminal{wait} statement, if any, i.e.\ it advances its pc;

\item[\normalfont\terminal{broadcast}] advances the pc of the executing
  thread and wakes up \emph{all} threads currently blocked at a
  \terminal{wait}.

\end{description}
\terminal{Wait} and release via \terminal{signal} or \terminal{broadcast}
are powerful synchronization mechanisms, allowing many threads to change
state at the same time. None of the above six statements change global
variables; only \terminal{start_thread} and \terminal{end_thread} change
the number of threads. \figureref{main example} (left) shows an example of
a BP with an assertion. A precise small-step operational semantics for
multi-threaded BPs is given in \ifthenelse{\arxivVersion}{\appendixref{Multi-Threaded Boolean
  Programs: Semantics}}{App. A of \cite{LW17b}}.

\subsection{From Boolean Programs to Thread Transition Systems}
\label{section: From Boolean Programs to Thread Transition Systems}

Transition systems are the input formalism for many exploration algorithms,
such as breadth-first search for reachability analysis, or the Karp-Miller
algorithm for deciding \emph{coverability} in infinite-state systems
\cite{KM69}. To apply these to BPs (and thus connect them, via predicate
abstraction, to software verification), the programs are typically
translated into transition systems.

Let Boolean program $\B$ be defined over sets of global and local variables
$\sharedvars$ and $\localvars$, respectively, and let $\shortrange 1
{\pc_{\max}}$ be the set of program locations.\footnote{We write
  $\shortrange l r$ compactly for $\{n \in \mathbb N: l \atm n \atm r\}$.}
We translate $\B$ into a finite-state \emph{thread transition system} (TTS)
$M=(S,R)$, over the state space $S = \{0,1\}^{|\sharedvars|} \times
\shortrange 1 {\pc_{\max}} \times \{0,1\}^{|\localvars|}$ and \emph{edges}
$R$.

Individual BP statements are translated into edges, as follows. A given
state $s \in S$ determines a program state $s_\B$ of $\B$ in a
straightforward way: $s$ encodes a valuation of all global variables (the
$\{0,1\}^{|\sharedvars|}$ part, the \emph{global state}), a program
counter, and a valuation of all local variables (the
$\{0,1\}^{|\localvars|}$ part, the \emph{local state}). Executing $\B$
on~$s_\B$
has several effects: first, it generally changes both the global variables,
and the local variables of the active thread (including the pc). These
changes result in a new state $t \in S$ again in a straightforward way,
defining an edge $(s,t) \in R$.

\begin{figure*}[htbp]
  \centerTwoOut{%
    \begin{minipage}{0.22\textwidth}
      \includegraphics[width=\textwidth]{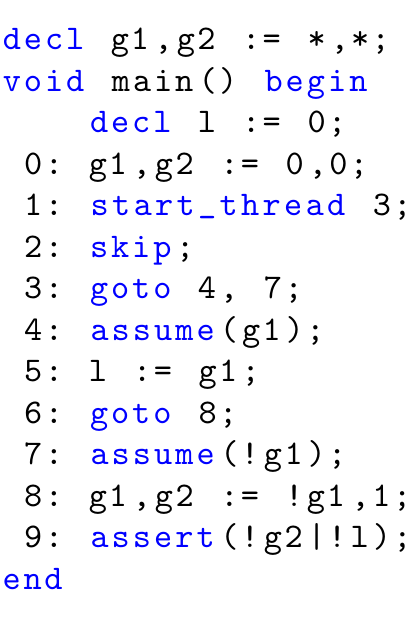}
    \end{minipage}
  }{%
    \begin{minipage}{0.76\textwidth}
      %\resizebox{\textwidth}{!}{\input{figures/tts.tex}}
      \includegraphics[width=\textwidth]{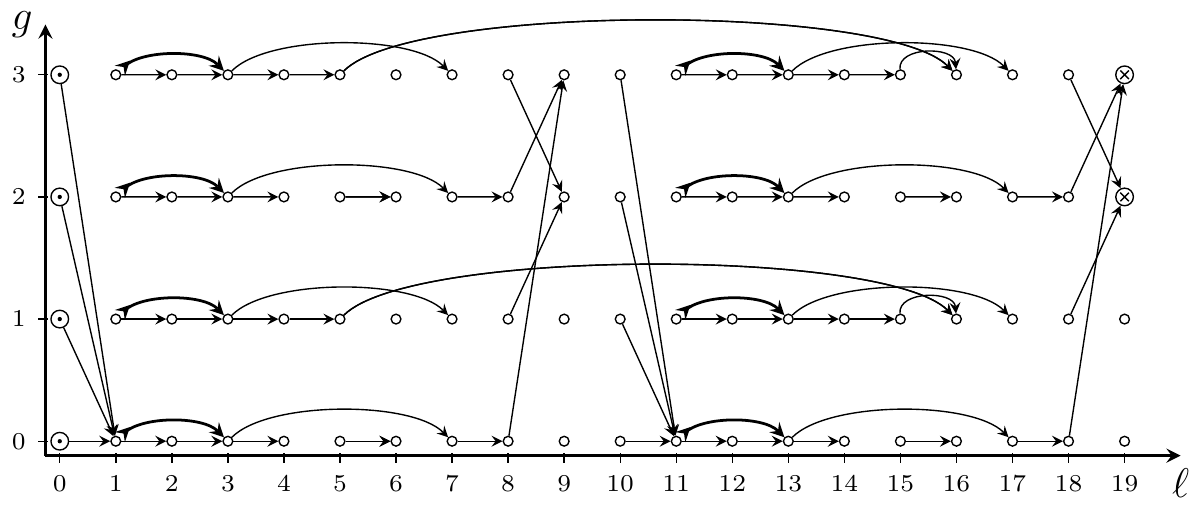}
    \end{minipage}}
  \vspace{-4pt}
  \caption{A Boolean program (left) and a possible translation into a TTS
    (right). Global variable valuation $(\texttt{g1},\texttt{g2})$ is
    encoded as state $g = 2 \times \texttt{g2} + \texttt{g1} \in \shortrange 0 3$. Similarly, local variable
    valuation $(\pc, \texttt{l})$ is encoded as state $\ell = 10 \times \texttt{l} + \pc \in
    \shortrange 0 {19}$. With this encoding, the four initial program
    states are shown as $\odot$\,, the two assertion failure states
    (satisfying $\pc = 9 \land \texttt{g2} = \texttt{l} = 1$) as $\otimes$.
   }
  \label{figure: main example}
\end{figure*}

Second, thread creation and termination, as well as signals and broadcasts,
typically have ``side effects'' that alter the thread count, or local
variables of passive threads. To capture such effects in the
(single-thread) data structure~$M$, each edge comes with a \emph{type}. It
is then left to the exploration algorithm, which has access to the current
system state, to fully implement transition semantics.
As an example, \figureref{main example} shows a BP and a translation into a
TTS. Symbol $\rightarrowtail$ marks edge $(0,1) \rightarrowtail (0,3)$ as a
thread creation edge. The semantics of thread creation
(\ifthenelse{\arxivVersion}{\appendixref{Multi-Threaded Boolean
  Programs: Semantics}}{App. A of \cite{LW17b}}) prescribes that
the active (creat\emph{ing}) thread moves on (to $\pc=2$); this is
reflected by an ordinary edge $(0,1) \rightarrow (0,2)$ in the
TTS. The creat\emph{ed} thread needs a start location, which is the pc
value of the BP state $(\texttt{g1},\texttt{g2},\pc,\texttt{l}) = (0,0,3,0)$
encoded by the target TTS state $(0,3)$ of the edge.
Other than above two types of edges shown in \figureref{main example},
there is one more type, denoted by $\rightsquigarrow$, used in the TTS
to characterize broadcasts.

The problem with such a translation from $\B$ to $M$ is of course the
potential blow-up: the nominal state space $S$ of $M$ is exponential in the
number of global and local variables.
This problem has long been known and has led to sophisticated
\emph{on-the-fly} temporal-logic model checkers such as \spin~\cite{GH97},
but also to ad-hoc re-implementations of specific exploration algorithms
\cite{BHKOWZ10,LW14}. In the rest of this paper we describe an API that
automates the construction of on-the-fly program state explorers.

\section{BP Analysis with JIT Translation: Overview}
\label{section: BP Analysis with JIT Translation: Overview}

We target \emph{exploration algorithms}, i.e.\ algorithms that operate
on a transition system representation of 
the given program and involve \emph{image computations}: given a system
state, they repeatedly compute some notion of successors or predecessors of
the state.\Ifthen{\arxivVersion}{\ Image computations are the primary operation in all
  kinds of state space exploration algorithms, such as elementary search or
  full finite-state model checking, but also infinite-state analyses such
  as the Karp-Miller algorithm for vector addition systems \cite{KM69} or
  the backward search coverability method for well quasi-ordered systems
  \cite{A10}. In the latter case, the notion of ``image'' is somewhat more
  complex: it proceeds backwards and may increase the dimension of a state.
  We will return to this algorithm later in the paper.} \figureref{standard
  vs. on-the-fly exploration} (left; ignore the boxes for now) shows a
schematic version of such algorithms. Input is a transition system $M$ and
some target state set $T$, such as a bad system state whose discovery would
indicate a reachable error in the system. The algorithm maintains a
worklist $W$ of states to be explored, typically initialized to the initial
or bad states of the system, depending on whether the search proceeds
forward or backward. It also maintains a set $X$ of explored states,
initially empty. The exploration proceeds by extracting an unexplored state
$w$ from $W$ and iterating through the set of states $w'$ in $w$'s image,
computed by $\StepMath$. If $w'$ is new,
we test whether it belongs to the target states $T$. If so, we report the
success of the search. The search terminates when no more unexplored states
exist (in $W$).

\newcommand{\changed}[1]{\fbox{#1}}

\begin{figure*}[htbp]
  \renewcommand{\algorithmicforall} {\plkeyword{for}} % to save space
  \begin{minipage}{60mm}
    \algsetup{indent=1.2em}
    \begin{algorithm}[H]
      \floatname{algorithm}{Scheme}
      \caption{$\textsc{Explore}(M,T)$}
      \begin{algorithmic}[1]
        \REQUIRE{\changed{transition system $M$}, target $T$}
        \STMT Initialize $W$ and $X$
        \WHILE{$\exists w \in W$} 
          \STMT $W := W \setminus \{w\}$
%          \STMT \changed{/\!/ \emph{\scriptsize this line intentionally left blank}} \label{line: blank}
          \FOREACH{\changed{$w' \in \StepMath(w)$}}                                  \label{line: step}
%            \STMT \changed{/\!/ \emph{\scriptsize ditto}}                            \label{line: blank 2}
%            \IF{$\NewMath{w'}{X}$}
              \IF{$w'$ not in $X$}
              \IF{$w'$ in $T$}
%             \IF{$\DecideMath(w',T)$}
                \RETURN ``found''
              \ENDIF
              \STMT merge $w'$ into $W$ and $X$                                      \label{line: update W and X}
            \ENDIF
          \ENDFOR
        \ENDWHILE
        \RETURN ``not found''                                                        \label{line: not found}
      \end{algorithmic}
      \label{scheme: standard}
    \end{algorithm}
  \end{minipage}
  \hfill
  \begin{minipage}{60mm}
    \algsetup{indent=1.2em}
    \begin{algorithm}[H]
      \floatname{algorithm}{Scheme}
      \caption{$\textsc{Explore\_Ijit}(\B,T)$}
      \begin{algorithmic}[1]
        \REQUIRE{\changed{Boolean Program $\B$}, target $T$}
        \STMT Initialize $W$ and $X$
        \WHILE{$\exists w% = (g, \vect{\ell})
          \in W$}% (\ell_1,\ldots,\ell_n)
          \STMT $W := W \setminus \{w\}$
%          \STMT \changed{$p := \sstops(w)$}                                            \label{line: sstops}
        % \FOREACH{$i = 1 \to |\vect{\ell}|$}
          \FOREACH{\changed{$w' \in \pstoss(\StepMath[\B](\sstops(w)))$}}            \label{line: otf image}
%           \IF{$\NewMath{w'}{X}$}
            \IF{$w'$ not in $X$}                                                     \label{line: new}
              \IF{$w'$ in $T$}                                                       \label{line: reachable}
%             \IF{$\DecideMath(w',T)$}
                \RETURN ``found''
              \ENDIF
              \STMT merge $w'$ into $W$ and $X$                                      \label{line: update W and X 2}
            \ENDIF
          \ENDFOR
        \ENDWHILE
        \RETURN ``not found''                                                        \label{line: not found 2}
      \end{algorithmic}
      \label{scheme: on-the-fly}
    \end{algorithm}
  \end{minipage}
  \caption{State exploration over a transition system (left) and a Boolean
    program (right). Lines~\lineref[]{new} and~\lineref[]{reachable} test
    whether $w'$ has not been explored and $w'$ is a target state,
    respectively. In a concrete algorithm these tests may involve more than
    set membership.}
  \label{figure: standard vs. on-the-fly exploration}
\end{figure*}

Now suppose the transition system $M$ is actually a translation of a
Boolean program $\B$, which we want to explore
directly, using the same algorithm scheme. One way to achieve that is to
change the data structure that \schemeref{standard} relies on: instead of
storing states to be explored as states of $M$, we store them as Boolean
program states, one entry per program variable\Ifthen{\arxivVersion}{\ (and perhaps per
  thread)}. Images are then computed by ``executing'' $\B$ in accordance
with $\B$'s execution model.

However, like with any data structure change in any non-trivial program,
the required effort is significant: all of $T$, $W$, $X$ must be changed,
and therefore virtually every line in a program that implements
\schemeref{standard}.
Re-implementing $\StepMath$ to operate on a Boolean program $\B$ is also
involved.
The whole change process is not only error-prone; it also creates an
entirely new implementation that needs to be maintained independently of
the one operating on $M$.

An alternative to this strategy is shown in \schemeref{on-the-fly} on the
right, which is almost identical to that on the left. States are stored as
transition system states of $M$ as before, but the input is now the Boolean
program $\B$. Since $M$ is no longer available, we cannot apply $M$'s
transition relation to compute images. However, since there is a one-to-one
correspondence between states of $\B$ and of $M$, we can compute images by
converting, using function $\sstops$, to $\B$'s state representation
\emph{just in time} for the image computation, and reverting the resulting
image states back to the system state format of $M$ (\lineref{otf image}).
Note that $\pstoss$ needs to operate on (and return) \emph{sets} of
states.

Operation $\StepMath[\B]$ computes images of an intermediate program state
\mbox{$p := \sstops(w)$}. Its implementation depends on the kind of image
computation performed by the algorithm: For standard forward exploration,
it can be computed by executing, from $p$, the statement of $\B$ pointed to
by the pc\Ifthen{\arxivVersion}{\ (of the active thread, in the multi-threaded case)}
encoded in $p$. For a backward exploration algorithm, $\StepMath[\B]$ is
more complicated: we need to identify statements \emph{leading to} the
current pc via $\B$'s control flow graph, and then
symbolically execute such statements backwards, e.g.\ via weakest
preconditions \cite{LW14}.\Ifthen{\arxivVersion}{\ This idea was presented in \cite{LW14}
  for the case of Abdulla's backward search algorithm~\cite{A10}.}

The API presented in this paper supplies an implementation of the $\B
\leftrightarrow M$ conversion functions ($\sstops,\pstoss$) and of various
common image operations applied to (multi-threaded) Boolean program states,
including backward statement execution for backward search algorithms. In
many cases, all the user needs to do is to replace the image operation in
their algorithm, as shown in \figureref{standard vs. on-the-fly exploration}
(boxes).

A minor runtime cost of using an algorithm according to
\schemeref{on-the-fly} is that the repeated conversion will take some time.
This time is linear in the number of Boolean program variables (and the
number of threads of the current system state, if multi-threaded). The
state conversion in either direction is a simple operation that can be
highly optimized. We will demonstrate in \sectionref{Case Study:
  Performance Benefits of IJIT} that the benefit of avoiding the explicit
construction of $M$ often far outweighs the conversion overhead.

\Ifthen{\arxivVersion}{We end this section by discussing desirable characteristics of
  algorithms that will benefit from using our API. We target exploration
  (search, model checking) algorithms for state transition systems
  (e.g.\ TTS) of Boolean programs. The term ``exploration'' here refers to
  the reliance of such algorithms on the computation of pre- and postimages
  of (sets of) states. The transition systems must relate to the Boolean
  programs in a way that there is a one-to-one correspondence between
  program states and system states. In particular, the systems cannot be
  (lossy) \emph{abstractions} of the Boolean programs; otherwise, a system
  state may not map to a unique program state, or vice versa.}

\section{The \ourapiplain\ Application Programming Interface} % 
\label{section: The API}

In this section we sketch usage and design of our API, named
\ourapi: \textbf Interface for \textbf Just-\textbf In-\textbf Time translation.
A detailed tutorial and documentation can be found in~\cite{IjitWebsite}.

\subsection{API Usage}
\label{section: API Usage}

We use a fictitious procedure \code{explore} to illustrate the use of our
API; see \figureref{an example of API usage} (left). The procedure explores
the state space of some transition system given as a TTS. It begins by
reading the TTS into a data structure called \code R (Line~5) and extracts
from \code R sets of initial and final states, respectively (Lines~7
and~8). The procedure then enters some kind of loop to explore the state
space represented by \code R, perhaps until no more unexplored states are
available (this is immaterial for our API). Crucial is that the loop body
will invoke an image operation on a state \code{tau} (Line~12), likely at
least once in each iteration. We assume \code R is nondeterministic, so
that the call returns a set of states, \code{Tau}.
\begin{figure}[htbp]
  %\resizebox{5.1cm}{!}{
    \begin{minipage}{0.41\textwidth}
\begin{lstlisting}[frame=trBL,
      commentstyle=\color{black!50!green!70}, 
      morekeywords={set},
      framexleftmargin=2pt,        
      numbers=right,              
      numbersep=10pt, 
      numberstyle=\tiny\rmfamily]
// user's headers, namespace, etc.

void explore() {
  // ...
  auto R = read_file("filename.tts")1

  set<state> I = R.init();
  set<state> F = R.final();
  state tau;
  while (...) { // state exploration
    tau = ... ; // an unexplored state
    set<state> Tau = image(tau);

    ...
  } // end while
}
\end{lstlisting}
    \end{minipage}%}
  \hfill
  %\resizebox{7.5cm}{!}{
    \begin{minipage}{0.51\textwidth}
\begin{lstlisting}[frame=trBL, 
      numbers=none,
      commentstyle=\color{black!50!green!70}, 
      morekeywords={set}, 
      framexleftmargin=2pt]
#include "ijit.hh"
using namespace ijit;
void explore_jit() {
  // ...
  auto P = @\hilight{parser::parse("filename.bp", mode::POST)}@;
  @\hilightg{converter c;}@
  set<state> I = @\hilightr{c.convert}@(P.init());
  set<state> F = @\hilightr{c.convert}@(P.final());
  state tau;
  while (...) {
    tau = ... ;
    set<state> Tau = @\hilightr{c.convert(}@
                @\hilightr{image(c.convert(tau), mode::POST))}@;
    ...
  }
}
\end{lstlisting}
    \end{minipage}%}
  \caption{An example illustrating the usage of \ourapi. Left: a fictitious
    state space exploration procedure. Right: the just-in-time version
    obtained using \ourapi. Line numbers in the middle; highlighted code
    shows places that have changed from the original
    version.}
  \label{figure: an example of API usage}
\end{figure}

\figureref{an example of API usage} (right) highlights (in \hilightcolor)
the changes the programmer needs to make to have procedure \code{explore}
operate on a Boolean program; we call the resulting procedure
\code{explore\_jit}. We explain these changes in the following.
\begin{enumerate}[$\bullet$]

\item Instead of reading a TTS, we now read a Boolean program as input
  (Line~5). This is done using a parser supplied by \ourapi. Procedure
  \code{parse} has two arguments: the name of input file, and the parser's
  direction mode: \code{POST} will cause the parser to generate code for
  subsequent forward-directed analysis (via postimages). Mode \code{PREV}
  does the analogous for backward analysis;
  a mode of \code{BOTH} will generate code for both. The parser also offers
  functionality to return sets \code I and \code F of initial and final
  program states, extracted from the initial variable declarations and
  assertions in the~BP, respectively.

\item The conversion between different state representation formats,
  explained below, is done via methods of a class \code{converter}. The
  user needs to instantiate this class before any conversion
  methods of the API can be called (Line 6).

\item Conversion between state representation formats happens in several
  places: to convert the initial and final Boolean program state sets into
  TTS state sets (Lines~7 and~8), and in the image computations. If the
  algorithm implemented by procedure \code{explore} operates on TTS as
  defined in \sectionref{Boolean Programs and Thread-Transition Systems},
  the JIT version of the procedure can be implemented using conversion
  functions supplied by the API (Line~12): the current (unexplored) TTS
  state \code{tau} is converted into a BP state, followed by a Boolean
  program image computation using the given direction mode, followed by a
  back-conversion into a set of TTS states.\Ifthen{\arxivVersion}{\ The API's
    \code{image} function by default returns a set of states.}

  If the API's conversion functions cannot be used, users must supply
  their own functions.
  To reduce the programming burden, the API provides an inheritance
  interface that allows defining conversion functions via specialization.
  Users are free to define stand-alone conversions.

\end{enumerate}

\subsection{API Design}
\label{section: API Design}

API \ourapi~is implemented in C++. A schematic overview is shown in
\figureref{schematic overview}.
\begin{figure}[htbp]
  \centering
  \includegraphics[scale=0.55]{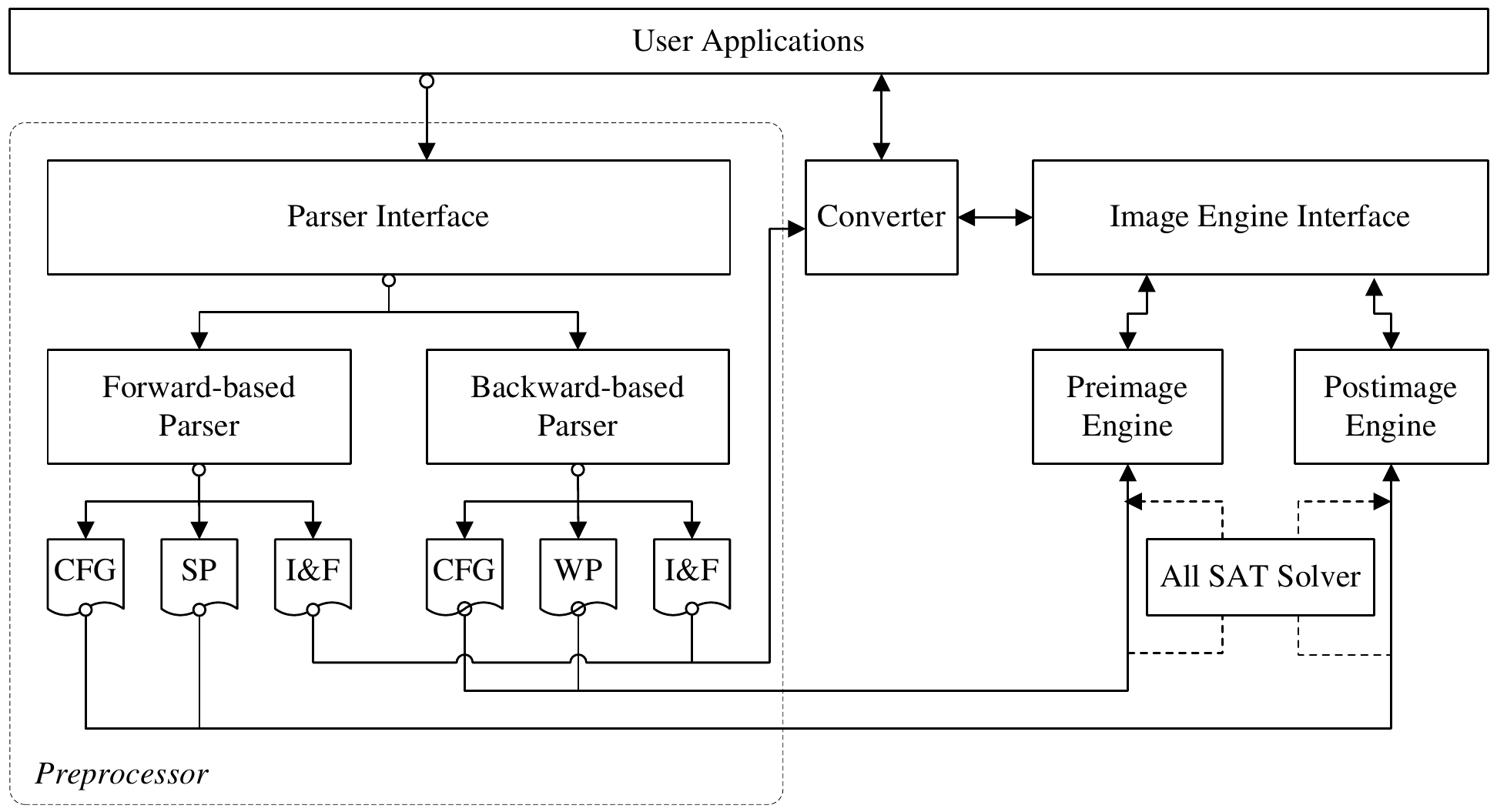}
  \caption{Schematic overview of \ourapi. The preprocessor part is
    usually called only once. CFG: control flow graph; SP/WP:
    strongest postcondition/weakest precondition; I/F: the set of
    initial/final states}
  \label{figure: schematic overview}
\end{figure}

\paragraph*{Parser.} The main purpose of the parser is to process the input
BP and populate the data structures to be used in image computations. These
include the program's control flow graph, and pre- and postcondition
expressions for pre- and postimage computations,
respectively.\Ifthen{\arxivVersion}{\ More precisely, assuming \code{mode::POST}
  directive, procedure \code{parse} causes the parser to generate strongest
  postcondition expressions formalizing the semantics of program statements
  for subsequent forward-directed analysis (via postimages), similarly for
  the other direction mode.} The parser also extracts initial and final
state information, the latter by collecting all states violating any of
assertions in the Boolean program.

\paragraph*{Converter.} The converter provides an adapter between system
states and program states.
In our design, the converter is an abstract C++ class with default
implementations of conversion functions.
If desired or necessary, users can either inherit the abstract class and
override the default implementation, or write a stand-alone converter from
scratch.

\paragraph*{Image Engine.} At the core of our API are the engines
to compute the preimage or postimage
of a given Boolean program state. These routines make use of the control
flow graph obtained by the parser, especially for preimages, in order to
determine the set of statements that can lead to the current
$\pc$\Ifthen{\arxivVersion}{\ (there can be several, e.g.\ due to the presence of
  \code{goto}s)}. Once the statement to be executed forward or backward has
been determined, the statement's semantics determines the effect on the
program data. The semantics is given as a set of first-order predicates
expressing strongest post- or weakest preconditions. To perform image
computations, the engine instantiates these formulas with the current-state
valuations of the program variables. It then invokes an All-SAT
solver to obtain
the pre- or post\-images as satisfiable assignments.

\paragraph*{All-SAT Solver.} The All-SAT solver used in image
computations is not based upon a state-of-the-art SAT solver, which would
require CNF conversion. Instead we found it to be more efficient to
simply build a custom SAT solver that enumerates solutions. Note that
input formulas to the solver formalize Boolean program statements and
thus tend to be very short.

\section{Case Study: Performance Benefits of \ourapiplain}
\label{section: Case Study: Performance Benefits of IJIT}

We evaluate the benefit of our API on a number of diverse benchmark
algorithms. All are designed to operate on thread-transition systems (TTS)
for either a fixed or an unbounded number of threads; we wish to apply them
to multi-threaded Boolean programs directly. For each algorithm, we compare
the performance of three versions: (i)~the \emph{TTS version}, which is the
original version, but prefixed by an input translation from BPs into TTS;
(ii)~the \emph{BP version}, which is a manual and \emph{optimized}
re-implementation where the internal state data structure has been changed
to BP states; and finally (iii) the \emph{JIT version}, which employs our
API. We expect a performance ranking of the form
\[
  \mbox{\emph{BP version}} \wbox{$<$} \mbox{\emph{JIT version}} \wbox{$\ll$} \mbox{\emph{TTS version}}
\]
where ``$<$'' (``$\ll$'') means ``(much) faster''. In particular, the
hand-crafted BP version makes repeated conversion between state
representations unnecessary and can therefore be considered the gold
standard for efficiency. We hope the automated JIT version of the algorithm
to perform nearly as well.\Ifthen{\arxivVersion}{\footnote{We note that, while not its
    purpose, our API is also useful for building the BP version of the
    algorithm, if so desired: it provides the parser and image operations
    for Boolean programs, as well as internal data structures for program
    state representation.}}

\subsection{Benchmark Algorithms}
\label{section: Benchmark Algorithms}

We sketch the purpose and basic concepts of four diverse
algorithms used in our case study; more details are provided 
in \ifthenelse{\arxivVersion}{\appendixref{Benchmark Algorithms in
    Appendix}}{App. B of \cite{LW17b}}. 
The algorithms
cover the spectrum of finite- and infinite-state searches, and of forward
and backward explorations.

\subsubsection{Cutoff Detection via Finite-State Search (\ECUT)~\cite{KKW10}.}
\ECUT\ implements \emph{dynamic cutoff detection} for parameterized
thread transition systems.  
A \emph{cutoff} point is a number $n_0$ of threads that are sufficient
to reach all reachable thread states. 
The core procedure of \ECUT\ is a (multi-threaded but) finite-state
search, BFS style. 
The TTS version of \ECUT\ can 
be transformed into the JIT version without any programming beyond the few
changes discussed in \sectionref{The API}.

\subsubsection{Karp-Miller Procedure~\cite{KM69}.}
We experiment with two variants of this classic procedure; both are in use in
unbounded-thread program verification:
\begin{enumerate}[(1)]
\item \KM\ decides the reachability of a specific target state $t$:
  it stops when a state covering $t$ has been encountered;

\item \AKM\ (``All-\KM'') builds the complete coverability tree,
  i.e.\ it runs \KM\ until a fixpoint is reached.
\end{enumerate}

\subsubsection{WQOS Backward Search (BWS)~\cite{ACJT96,A10}.} This technique
is a sound and complete algorithm to decide coverability for \emph{well
  quasi-ordered systems} (WQOS), a~broad family of transition systems that
subsumes replicated Boolean programs, Petri nets, VASS, and many more. Note
that BWS is a backward exploration. In contrast, the previous three
algorithms explore forward.

\subsection{Case Study}
\label{section: Case Study}

\paragraph*{Experimental Setup.} We compare the impact of our API on the
efficiency of the four algorithms described in \sectionref{Benchmark
  Algorithms}. For each algorithm $A \in \{\textsc{Ecut}, \textsc{Km},
\textsc{Akm},$ \format $\textsc{Bws}\}$, we compare three different
versions: (1) the TTS version --- named \version A {\tts}; (2) the
\apiotf\ version obtained using our API --- named \version A {\apiotf}; and
(3) the hand-implemented Boolean program version --- named \version A
                 {\apibp}.

We perform the comparison using a collection of Boolean programs obtained
via predicate abstraction from 30 concurrent \C\ programs\Ifthen{\arxivVersion}{\ (our
  benchmark algorithms are intended for concurrent program analysis)}. The
\C\ programs are detailed in \tableref{benchmarks}.
We use \SATABS\ \cite{CKSY05} to construct the BPs from these programs. The
BPs are also concurrent; threads execute the same Boolean procedure. In
most cases, the same \C\ source program generates several BPs (since
\SATABS\ goes through several abstract-verify-refine iterations). In the
end we obtained 155 BPs for the 30 \C\ programs. For the TTS version of
each algorithm,
we use \SATABS\ to generate the TTS from the Boolean program (option
\texttt{--build-tts}; this is where the input format explosion inevitably
happens).
\begin{table*}[t]
 \caption{Benchmark statistics: $\mathit{GV}$/$\mathit{LV}$/$\LOC$ = \# of
   global/local \C\ program variables/lines of code;
   $\numsharedvars$/$\numlocalvars$/$|\mathit{PC}|$/\emph{Its.} = \# of
   global/local Boolean variables/program counters/CEGAR iterations;
   $\numshares$/$\numlocals$/$|R|$ = \# of global/local states/transitions
   in TTS; \emph{Safe?} = $\YES$ : program safe; $|\cdot|$ represents the
   median of the feature across different BP/TTS resulting from the same
   \C\ program. Note that often $\numshares > 2^{|\sharedvars|}$, due to
   auxiliary states used by \SATABS\ in the BP $\rightarrow$ TTS
   translation}
 \vspace{2mm}
  \label{table: benchmarks}
  \centering
\resizebox{6.0cm}{!}{
  % \begin{minipage}{0.7\linewidth}
    \centering
    \noindent
    \begin{tabular}{@{}lrrrc@{}rrrrc@{}rrr@{}c@{}}
     \hline
        \multirow{2}{*}{ID/Program} & \multicolumn{3}{c}{C Program} & \multirow{2}{*}{\ \ } & \multicolumn{4}{c}{BP} 
       & \multirow{2}{*}{\ \ } & \multicolumn{3}{c}{TTS}  & \multirow{2}{*}{~\emph{Safe?}} \\
       \cline{2-4}\cline{6-9}\cline{11-13}\noalign{\smallskip} 
       & \multicolumn{1}{@{}c@{}}{$\mathit{GV}$} & \multicolumn{1}{c@{}}{$\mathit{LV}$} & \multicolumn{1}{c}{$\LOC$} 
       & & \multicolumn{1}{@{}c@{}}{$\numsharedvars$} & \multicolumn{1}{c@{}}{$\numlocalvars$} & \multicolumn{1}{c@{}}{$|\mathit{PC}|$}  &  \multicolumn{1}{c@{}}{\emph{Its.}}
       & & \multicolumn{1}{@{}c@{}}{$\numshares$} & \multicolumn{1}{c@{}}{$\numlocals$} & \multicolumn{1}{c}{$|R|$} & \\
      \hline
      01/\textsc{Increm-l}    & $2$ & $1$ & $46$ & & $3$ & $1$ & $40$ & $2$ & & $33$ & $71$ & $688$ & $\YES$\\
      02/\textsc{Increm-c}    & $1$ & $3$ & $57$ & & $0$ & $4$ & $35$ & $4$ & & $5$ & $449$ & $784$ & $\YES$ \\
      03/\textsc{Prng-l}    & $2$ & $4$ & $63$ & & $2$ & $3$ & $45$ & $2$ & & $17$ & $265$ & $1488$ & $\YES$ \\
      04/\textsc{Prng-c}    & $1$ & $5$ & $95$ & & $0$ & $5$ & $48$ & $2$ & & $5$ & $993$ & $1760$ & $\YES$ \\
      05/\textsc{FindMax-l}    & $3$ & $3$ & $59$ & & $1$ & $0$ & $43$ & $2$ & & $9$ & $25$ & $57$ & $\YES$ \\
      06/\textsc{FindMax-c}    & $2$ & $5$ & $79$ & & $0$ & $1$ & $48$ & $2$ & & $5$ & $59$ & $76$ & $\YES$ \\
      07/\textsc{MaxOpt-l}    & $3$ & $4$ & $69$ & & $1$ & $1$ & $48$ & $2$ & & $9$ & $63$ & $162$ & $\YES$ \\
      08/\textsc{MaxOpt-c}    & $2$ & $6$ & $86$ & & $0$ & $2$ & $53$ & $2$ & & $5$ & $137$ & $196$ & $\YES$ \\
      09/\textsc{Stack-l}     & $4$ & $2$ & $79$ & & $1$ & $3$ & $53$ & $3$ & & $9$ & $157$ & $360$ & $\YES$ \\
      10/\textsc{Stack-c}     & $3$ & $3$ & $89$ & & $3$ & $1$ & $54$ & $2$ & & $33$ & $81$ & $740$ & $\YES$ \\ 
      11/\textsc{Bs-Loop}     & $0$ & $6$ & $24$ & & $0$ & $7$ & $30$ & $1$ & & $65$ & $24$ & $448$ & $\NO$ \\
      12/\textsc{Cond}        & $1$ & $3$ & $56$ & & $0$ & $3$ & $29$ & $2$ & & $33$ & $25$ & $200$ & $\YES$ \\
      13/\textsc{Func-P}      & $2$ & $1$ & $67$ & & $2$ & $6$ & $32$ & $3$ & & $5$ & $3969$ & $9728$ & $\YES$ \\
      14/\textsc{S-Loop}      & $5$ & $0$ & $60$ & & $4$ & $0$ & $37$ & $20$ & & $5$ & $209$ & $296$ & $\YES$ \\
      15/\textsc{Pthread}     & $5$ & $0$ & ~$85$ & & $7$ & $0$ & ~$60$ & $5$ & & ~~~$17$ & ~$3329$ & ~$20608$ & $\NO$ \\
      \hline
    \end{tabular}                               
  % \end{minipage}
}
  \hfill
\resizebox{6.0cm}{!}{
  % \begin{minipage}{0.7\linewidth}
    \centering
    \noindent
    \begin{tabular}{@{}lrrrc@{}rrrrc@{}rrr@{}c@{}}
     \hline
        \multirow{2}{*}{ID/Program} & \multicolumn{3}{c}{C Program} & \multirow{2}{*}{\ \ } & \multicolumn{4}{c}{BP} 
       & \multirow{2}{*}{\ \ } & \multicolumn{3}{c}{TTS}  & \multirow{2}{*}{~\emph{Safe?}} \\
       \cline{2-4}\cline{6-9}\cline{11-13}\noalign{\smallskip} 
       & \multicolumn{1}{@{}c@{}}{$\mathit{GV}$} & \multicolumn{1}{c@{}}{$\mathit{LV}$} & \multicolumn{1}{c}{$\LOC$} 
       & & \multicolumn{1}{@{}c@{}}{$\numsharedvars$} & \multicolumn{1}{c@{}}{$\numlocalvars$} & \multicolumn{1}{c@{}}{$|\mathit{PC}|$}  &  \multicolumn{1}{c@{}}{\emph{Its.}}
       & & \multicolumn{1}{@{}c@{}}{$\numshares$} & \multicolumn{1}{c@{}}{$\numlocals$} & \multicolumn{1}{c@{}}{$|R|$} & \\
      \hline
      16/\textsc{Tas-Lock}    & $2$ & $2$ & $58$ & & $3$ & $1$ & $48$ & $2$ & & $16385$ & $54$ & $269488$ & $\YES$ \\
      17/\textsc{DbLock-1}    & $3$ & $0$ & $70$ & & $7$ & $1$ & $79$ & $10$ & & $513$ & $151$ & $20928$ & $\YES$ \\
      18/\textsc{DbLock-2}    & $3$ & $0$ & $73$ & & $6$ & $1$ & $47$ & $22$ & & $33$ & $71$ & $688$ & $\YES$ \\
      19/\textsc{DbLock-3}    & $3$ & $0$ & $66$ & & $4$ & $1$ & $73$ & $3$ & & $257$ & $67$ & $4976$ & $\YES$ \\
      20/\textsc{Ticket-hc}\hspace{3pt}   & $3$ & $1$ & $61$ & & $5$ & $1$ & $73$ & $5$ & & $257$ & $139$ & $10912$ & $\YES$ \\
      21/\textsc{Ticket-lo}   & $3$ & $1$ & $46$ & & $5$ & $1$ & $63$ & $5$ & & $65$ & $115$ & $2048$& $\YES$ \\
      22/\textsc{BSD-ak}      & $1$ & $7$ & $90$ & & $3$ & $1$ & $119$ & $2$ & & $33$ & $196$ & $1922$ & $\YES$ \\
      23/\textsc{BSD-ra}      & $2$ & $21$ & $87$ & & $3$ & $0$ & $138$ & $2$ & & $33$ & $107$ & $996$ & $\YES$ \\
      24/\textsc{NetBSD}      & $1$ & $28$ & $152$ & & $3$ & $1$ & $278$ & $3$ & & $33$ & $423$ & $4096$ & $\YES$ \\
      25/\textsc{Solaris}     & $1$ & $56$ & $122$ & & $5$ & $1$ & $182$ & $2$ & & $129$ & $283$ & $10847$ & $\YES$ \\
      26/\textsc{Boop}        & $5$ & $2$  & $89$ & & $5$ & $2$ & $61$ & $4$ & & $129$ & $213$ & $8064$ & $\NO$ \\
      27/\textsc{Qrcu-2}      & $7$ & $6$ & $120$ & & $3$ & $0$ & $129$ & $15$ & & $33$ & $103$ & $1001$ & $\YES$ \\
      28/\textsc{Qrcu-4}      & $8$ & $8$ & $182$ & & $5$ & $2$ & $275$ & $21$ & & $129$ & $873$ & $35024$ & $\YES$ \\
      29/\textsc{Unver-if}    & $2$ & $1$ & $25$ & & $4$ & $0$ & $53$ & $3$ & & $129$ & $95$ & $4096$ & $\YES$ \\
      30/\textsc{SpinLock}    & $2$ & $0$ & $37$ & & $3$ & $0$ & $47$ & $2$ & & $129$ & $79$ & $3584$ & $\YES$ \\
      \hline
    \end{tabular}                              
  % \end{minipage}
}                              
\end{table*}

For each benchmark, we consider verification of a safety property,
specified via an assertion that is pushed, during predicate abstraction,
from \C\ to the Boolean program. All experiments are performed on a 2.3 GHz
Intel Xeon machine with 64 GB memory, running 64-bit Linux. The timeout is
set to 30 minutes; the memory limit to 4 GB. All benchmarks and
implementations are available at~\cite{IjitWebsite}.

\renewcommand{\localCommand}{\protect\makebox[8mm][r]{$\blacktriangleright$\ }}
\begin{figure*}[htbp]
  \centering
  \includegraphics[scale=1.07,trim=0.5cm 0 -0.2cm 0.3cm]{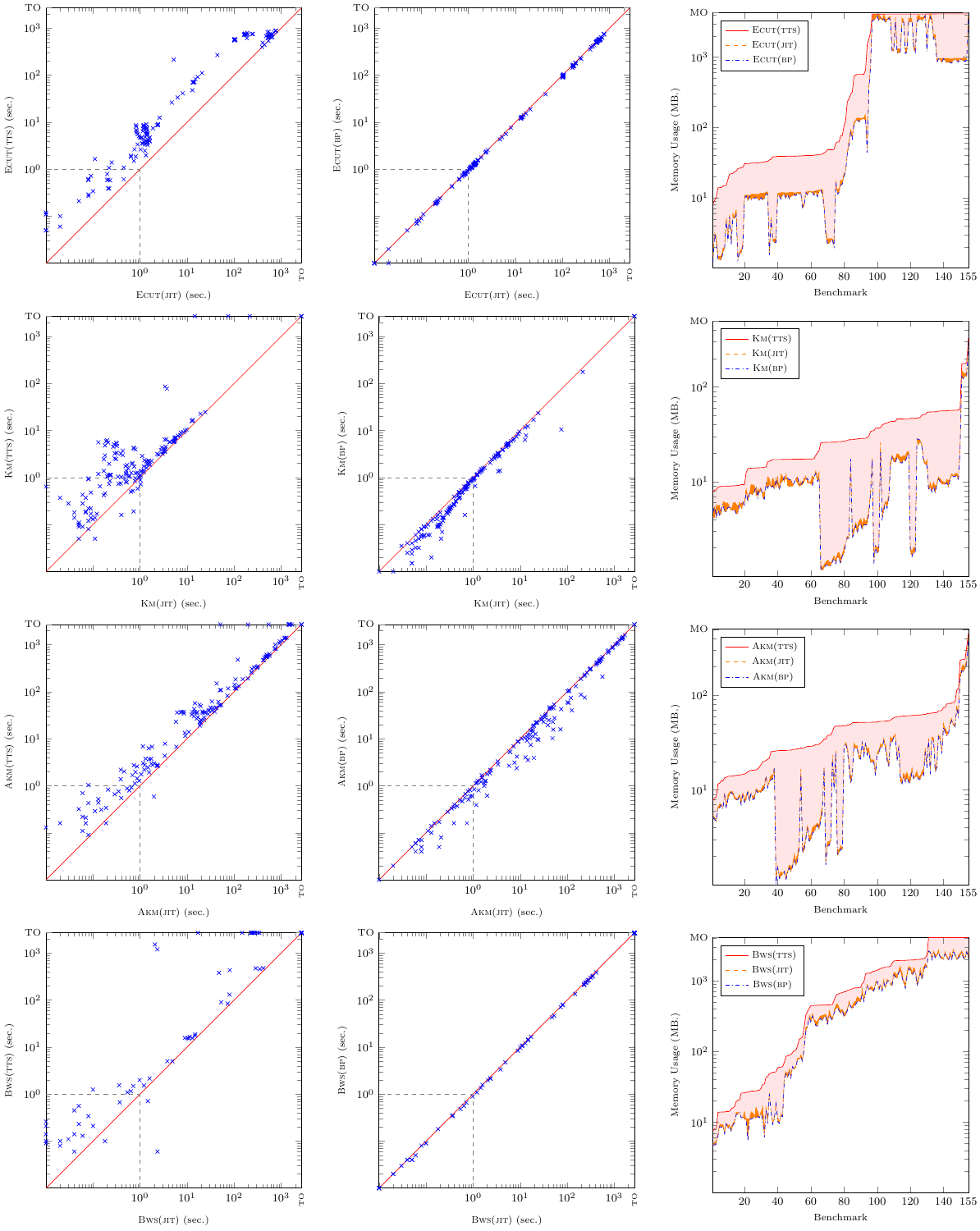}
  \vskip-4pt
  \caption{Performance impact of our API (TO: timeout, MO: memory out). \protect \\
    For $\mbox A \in \{\mbox{\textsc{Ecut}, \textsc{Km}, \textsc{Akm}, \textsc{Bws}}\}$: \protect \\
    \localCommand runtime comparison: left column: \version A {\tts}
    against \version A {\apiotf}; center: \version A {\apibp} against
    \version A {\apiotf}. Each dot = execution time on one example. Square
    in the lower left corner of each chart: runtime of less than 1 second
    for both algorithms, hence unreliable. \protect \\
    \localCommand memory usage comparison: right column: comparing memory
    usage across the three different versions. The plots are
    sorted by the memory usage of the TTS version of A. The shadowed
    areas show the difference. (Best viewed
    on-screen in color.)}
  \label{figure: experiment results}
\end{figure*}

\paragraph*{Results.} The results of our case study are shown in
\figureref{experiment results}. The first column shows, for the four
algorithms, the runtime comparison of the
\apiotf\ version\Ifthen{\arxivVersion}{\ obtained using \ourapi} (lower right in each
chart) against the original TTS version of the algorithm (upper left). The
log-scale charts clearly demonstrate the performance advantage ---
sometimes several orders of magnitude -- of not pre-translating
the input BP into a potentially large TTS.
%, but leaving this to the runtime.
In
many cases, runs that timed out in the TTS version can now be completed
within the 30mins limit. We point out that, while the conversion time BP
$\rightarrow$ TTS is included in the runtime for the TTS version, it is not
even to blame for the weaker TTS version performance: the conversion
usually takes a few seconds. What makes the TTS version slow is the
relatively large input TTS to the TTS-based algorithm.

The second column shows the runtime comparison of the
\apiotf\ version\Ifthen{\arxivVersion}{\ obtained using \ourapi} (lower right in each
chart) against the hand-implemented \apibp\ version of the algorithm (upper
left). Here the expectation is the opposite: we would like to get as close
to the diagonal as possible. This is achieved in all four cases to a
satisfactory degree.
% It is striking that
For the backward search algorithm, the comparison is more favorable for
\apiotf\ than for the two KM-based algorithms, with a performance nearly
indistinguishable from that of the \apibp\ version. This can be attributed
to the fact that \BWS\ overall takes more time than the forward search
implemented in \KM, since backward exploration faces more nondeterminism
and in general visits a larger number of configurations. The relative
overhead of state representation conversion is thus smaller.

The third column shows that the memory consumption of the \apiotf\ and
\apibp\ versions of each algorithm are very similar, and both are vastly
below that of the \tts\ version. This reflects in part the fact that the
\tts\ version needs to store the (relatively large) generated TTS in
memory. More relevant, however, is the fact that the TTS contains many
redundant (since unreachable) transitions --- their absence is the very
advantage of on-the-fly exploration techniques. Such redundant transitions
translate into a large number of redundant configurations explored by the
TTS version of the algorithm.

\section{Related Work}

Promoted by the success of predicate-abstraction based tools such as
\SLAM~\cite{BR02} and \SATABS~\cite{CKSY05},
Boolean programs are widely used in
verification. Accordingly, extensive research has been done on their
analysis, leading to a series of efficient algorithms, e.g., recursive
state machines~\cite{ABEGRY05}, and the symbolic verifiers
\textsc{Bebop}~\cite{BR00}, \textsc{Moped}~\cite{EHRS00,ES01},
\textsc{Boppo}~\cite{CKS05}, and \textsc{Getafix}~\cite{LPP09}.
Most of the above approaches use BDDs as symbolic representation, which do
not lend themselves to an efficient on-the-fly model construction.

In contrast, explicit-state model checking techniques often construct the
state space of the program they are exploring on the fly.
A prominent tool that pioneered this strategy is the explicit-state model
checker \spin~\cite{GH97}.
Another notable explicit-state on-the-fly model
checker is Java PathFinder~\cite{VHBPL03}, which takes $\mathrm{Java^{TM}}$
bytecode and analyses all possible paths through the program, checking for
deadlocks, assertion violations, etc.

Solutions addressing the translation blow-up in connection with (more
complex) unbounded-thread verification techniques are rare. While these
techniques have been applied to program analysis, the application is
typically preceded by an up-front translation of the program into an
explicit transition system \cite{DRB02,KKW10,KKW14a}. For Boolean programs
generated via predicate abstraction, this only works for small local state
spaces, for example when the number of predicates is small. When going
through several iterations of the predicate abstraction CEGAR loop, in
contrast, the number of Boolean program variables quickly becomes large.

On-the-fly techniques for unbounded-thread algorithms applied to Boolean
programs are given in tools by Basler et al.~\cite{BHKOWZ10}, and by Liu et
al.~\cite{LW14}. Both are re-implementations of the algorithms they are
targeting, which is the Karp-Miller procedure for VASS in the former case,
and the backward search algorithm for broadcast Petri nets in the latter.
Both demonstrate the benefits of exploring BPs directly, but they do not
come for free: the re-implementation is low-level work involving tricky
data structure changes, affecting the very foundation of the
implementation. In fact, the Karp-Miller implementation in~\cite{BHKOWZ10}
generated runtime errors on some of our benchmarks, so we excluded it from
our case study.

\section{Summary}

The problem of the blow-up between programs and transition systems that
describe the programs' semantics and are often used in exploration
algorithms is well known.\Ifthen{\arxivVersion}{\ It is a severe problem: beyond plain
  finite-state model checking, algorithms for infinite-state reachability
  analysis often already have high complexity (such as EXPSPACE for Petri
  net coverability \cite{CLM76,R78}).} Translating a program into an
explicit transition system undermines the practical runtime performance of
these algorithms, and thus diminishes their value. This problem has been
addressed in an ad-hoc way, by re-implementing these algorithms into ones
operating on programs. This process is painful and prone to programming
errors, to which we attribute the fact the input translation cost is often
grudgingly accepted.

In this paper we have introduced an API that largely automates the required
transformations. In the best case, programmers mostly need to add calls to
an API-provided \code{convert} method to (usually few) places in the code
where images are computed. In the worst case, programmers have to supply
this conversion method. We have demonstrated the huge impact of the use of
the API on various algorithms that rely on an up-front BP $\rightarrow$ TTS
translation. We have also compared the performance of the \apiotf\ version
to the version re-implemented by hand that operates entirely on Boolean
programs, and found nearly no performance difference to this gold-standard
implementation.

We have presented our API with dedicated support for algorithms that
operate on Boolean programs and thread-transition systems, due to their
popularity in, and significance for, software verification. Given proper
state representation conversion functions, we believe our API to be able to
bridge the gap between other types of modeling languages, such as Boolean
programs and Petri nets. We leave implementing, and experimenting with,
such extensions for the future.

%=====================================Reference================================%

% \newpage

\bibliographystyle{include/splncs03}
\bibliography{include/paper}

%=====================================Appendix=================================%
\Ifthen{\arxivVersion}{
\newpage

\appendix

\section*{Appendix}

\section{Multi-Threaded Boolean Programs: Semantics}
\label{appendix: Multi-Threaded Boolean Programs: Semantics}

We describe a transition system that formalizes the behavior of a possibly
unbounded number of threads concurrently running a Boolean program $\B$.
Let $\B$ be defined over sets of global and local Boolean variables
$\sharedvars$ and $\localvars$, respectively, and let $\shortrange 1
{\pc_{\max}}$ be the set of program locations.\footnote{We write
  $\shortrange l r$ compactly for $\{n \in \mathbb N: l \atm z \atm r\}$.}
$\B$ gives rise to a transition system $\Minf$ as follows. The states of
$\Minf$ have the form $(g,\ell_1,\ldots,\ell_n)$, where $g$, the
\emph{global state}, is a valuation of the global variables of $\B$. Symbol
$\ell_i$ denotes the \emph{local state of thread $i$}, a valuation of the
pc and the local variables of $\B$. A pair $\threadstate g {\ell_i}$ is thus
a program state of $\B$. We write $g.v$ ($\ell_i.v$) for the value of
global (local) variable $v$ in global (local) state $g$ ($\ell_i$),
$\ell_i.\pc$ for thread $i$'s current pc value, and $\stmt(\pc)$ for the
statement at the given pc. Finally, $n$ is a positive integer, intuitively
the number of threads currently running. The state space of $\Minf$ is
therefore the (infinite) set
\[
  \Sinf = \{0,1\}^{|\sharedvars|} \times \Union_{n=1}^\infty \left(\shortrange 1 {\pc_{\max}} \times \{0,1\}^{|\localvars|}\right)^n \ .
\]

Transitions
% in $\Minf$ % commented out to avoid ugly linebreak
have the form $(g,\ell_1,\ldots,\ell_n) \wbox[3pt]{$\transrelinf$}
(g',\ell'_1,$ $\ldots,\ell'_{n'})$ such that one of the following
conditions holds. To keep the description manageable, we use the convention
that \emph{variables not explicitly mentioned in the condition are
  unchanged in the transition} (this applies also to $n$).
\begin{description}

\item[%single-
thread transition:]
% $n' = n$ and
  there exists $i \in \shortrange 1 n$ such that
  $\stmt(\ell_i.\pc)$ is a \nonterminal{seqstmt}; executing it
  \emph{atomically} from the variable valuation given by $(g,\ell_i)$
  results in the variable valuation given by $(g',\ell'_i)$.

%, and

%  \item for $j \in \shortrange 1 n \setminus \{i\}$, $\ell'_j = \ell_j$.

%  \end{enumerate}

\item[thread creation:] $n' = n + 1$,
% $g' = g$
  and there exists $i \in \shortrange 1 n$ such that:
  \begin{enumerate} \alphaenumi

  \item $\stmt(\ell_i.\pc) = \code{start_thread $x$}$ for some $x$,

  \item $\ell'_i.\pc = \ell_i.\pc + 1$,
% and $\ell'_i.v = \ell_i.v$ for $v \in \localvars$,

  \item $\ell'_{n'}.\pc = x$ and $\ell'_{n'}.v = \ell_i.v$ for $v \in
    \localvars$.

%, and

%  \item for all $j \in \shortrange 1 n \setminus \{i\}$, $\ell'_j = \ell_j$.

  \end{enumerate}

\item[thread termination:] $n \atl 1$, $n' = n - 1$,
% $g' = g$,
  and there exists $i \in \shortrange 1 n$ such that
  \begin{enumerate} \alphaenumi

  \item $\stmt(\ell_i.\pc) = \code{end_thread}$,

  \item for all $j \in \shortrange 1 {(i-1)}$, $\ell'_j = \ell_j$, and

  \item for all $j \in \shortrange i {n'}$, $\ell'_j = \ell_{j+1}$ (left-shift).

  \end{enumerate}

\item[signal:]
% $n' = n$, $g' = g$ and
  there exists $i \in \shortrange 1 n$ such that
  \begin{enumerate} \alphaenumi

  \item $\stmt(\ell_i.\pc) = \code{signal}$,

  \item $\ell'_i.\pc = \ell_i.\pc + 1$,
% and $\ell'_i.v = \ell_i.v$ for $v \in \localvars$,
    and

  \item for all $j \in \shortrange 1 n \setminus \{i\}$, $\stmt(\ell_j.pc)
    \not= \code{wait}$ \hfill \emphasize{OR}
% $\ell'_j.pc = \ell_j.pc$ and $\ell'_j.v = \ell_j.v$ for $v \in
% \localvars$,
    there exists $j \in \shortrange 1 n \setminus \{i\}$ such that
    $\stmt(\ell_j.\pc) = \code{wait}$ and $\ell'_j.pc = \ell_j.pc + 1$.

%, $\ell'_j.v = \ell_j.v$ for $v \in \localvars$, and for $k
%\not\in\{i,j\}$, $\ell'_k.pc = \ell_k.pc$, $\ell'_k.v = \ell_k.v$ for $v \in \localvars$.

  \end{enumerate}

\item[broadcast:]
% $n' = n$, $g' = g$ and
  there exists $i \in \shortrange 1 n$ such that
  \begin{enumerate} \alphaenumi

  \item $\stmt(\ell_i.\pc) = \code{broadcast}$,

  \item $\ell'_i.\pc = \ell_i.\pc + 1$, and
% $\ell'_i.v = \ell_i.v$ for $v \in \localvars$,

  \item for all $j \in \shortrange 1 n \setminus \{i\}$ such that
    $\stmt(\ell_j.pc) = \code{wait}$, $\ell'_j.pc = \ell_j.pc + 1$.
% and $\ell'_j.v = \ell_j.v$ for $v \in \localvars$, and

%% \item for all $j \in \shortrange 1 n \setminus \{i\}$ such that the
%%   statement at $\ell_j.\pc$ is \emph{not} \terminal{wait}, $\ell'_j.pc =
%%   \ell_j.pc$ and $\ell'_j.v = \ell_j.v$ for $v \in \localvars$.

  \end{enumerate}

\end{description}
Note in particular that a \emphasize{thread transition} changes the
variables of at most one thread, that the other four transition types do
not change the global state, and that only \emphasize{thread creation} and
\emphasize{termination} change $n$. In each case, thread $i$ is called
\emph{active}, the others \emph{passive}. We omit the precise formalization
of \terminal{atomic} blocks, which is straightforward. The initial states
of $\Minf$ are obtained by setting $n$ to the initial number of threads
(typically 1), and $g$ and $\ell_i$ --- for any initially existing thread
$i$ --- according to the initial conditions for $\B$; in particular,
\format\linebreak\mbox{$\ell_i.\pc = 1$}.

\section{Benchmark Algorithms}
\label{appendix: Benchmark Algorithms in Appendix}

We give more details on the reference algorithms used in our Case Study
(\sectionref{Case Study: Performance Benefits of IJIT}).
%% A detailed exposition is beyond the scope of this paper; references
%% are provided. 

\subsubsection{Cutoff Detection via Finite-State Search.}

%% Thread-state cutoff detection is a significant verification approach for
%% parameterized programs, as there is a strong belief (backed by empirical
%% evidence) that parameterized systems often enjoy a small model property.
%% Unfortunately, most of above approaches are reasoning upon transition
%% systems.

% \thomas{before I shorten this: what do our experiments have to do with
%   cutoff detection? don't we simply run a fixed-thread exploration? why
%   explain cutoffs??}

\ECUT~\cite{KKW10} implements \emph{dynamic cutoff detection} for
parameterized thread transition systems. In such systems, the number $n$ of
threads is a parameter fixed at the outset; thread
creation is not considered in~\cite{KKW10}. Due to a monotonicity property, the number
$\#R$ of reachable thread states (combination of global and local
  state of the input TTS) can only grow as $n$ increases. Since, on the
other hand, the number of thread states is finite and thus $\#R$ is finite,
this growth must eventually end at a \emph{cutoff} point: a number $n_0$ of
threads that are sufficient to reach all reachable thread states.

\ECUT\ attempts to detect $n_0$ by gradually increasing $n$ and checking a
certain condition that implies the cutoff has been reached. (This method is
incomplete. \cite{KKW10} considers a rather brute-force extension to make it complete, which is not
considered in the present case study) Analyzing
the $n$-thread replicated instance of the TTS is a (multi-threaded but)
finite-state search problem. \ECUT\ performs it using a straightforward
forward search, BFS style. The TTS version of \ECUT\ can
be transformed into the JIT version without any programming beyond the few
changes discussed in \sectionref{The API}.
%% Deviating from the implementation in~\cite{KKW10}, we extended \ECUT\ to
%% deal with thread creation, which is ubiquitous in concurrent
%% programs.\thomas{\ I propose to omit the preceding sentence, as it will
%%   cause confusion with the ``parameterized'' setting. I also don't see why
%%   this is important}

\subsubsection{Karp-Miller Procedure (KM).} Assertion checking for
unbounded-thread Boolean programs can be reduced to a vector addition
system (VASS) coverability problem: can we reach a system state that
``covers'' a given target state? The covers relation, $s \covers
  t$, states that for each integer vector component, the value stored in
  $s$ is at least the value stored in $t$. The relationship with replicated
  Boolean programs is that such programs can be equivalently represented
  using counter vectors, one counter for each local program state a thread
  can be in. A ``bad'' state is then for example one where at least one
  thread resides in a local state that violates some assertion. Its
  reachability can be expressed as a coverability problem. Coverability
can be solved (among many techniques) using the classical
Karp-Miller procedure \cite{KM69}. It constructs, in finite time, a rooted
tree $T$ that compactly represents the generally infinite set of covered
configurations of a VASS. Each node of $T$ is labeled with a pair
  consisting of a shared state and a vector over $\mathbb{N} \union
  \{\omega\}$~; symbol $\omega$ intuitively represents an unbounded number
  of threads in the local state denoted by the corresponding index in the
  vector. The introduction of $\omega$ counter values permits acceleration
  of this infinite-state search algorithm and, in the end, guarantees
  termination.

We evaluate our API on two variants of Karp-Miller that are in use in
unbounded-thread program verification:
\begin{enumerate}[1.]
\item \KM\ decides the reachability of a specific target state $t$:
  it constructs $T$ as described above but stops when a state
  covering $t$ has been encountered;

\item \AKM\ (``All-\KM'') builds the complete coverability tree $T$,
  i.e.\ it runs \KM\ until a fixpoint is reached.
\end{enumerate}
Construction of $T$ is based on a forward exploration equipped with the
acceleration step mentioned above, which introduces $\omega$'s into a node
label. The test leading to acceleration can be performed on the system
stated obtained \emph{after} the back-conversion of the encountered BP
state; the same test as employed in the TTS version of \KM\ can be
used. Our API also provides a slightly more efficient
  acceleration test that operates directly on the encountered BP state
  immediately after the image has been computed.

\subsubsection{WQOS Backward Search (BWS).} This algorithm by Abdulla et al.
\cite{ACJT96,A10} is a sound and complete algorithm to decide coverability
for \emph{well quasi-ordered systems} (WQOS), a broad family of transition
systems that subsumes replicated Boolean programs, Petri nets, VASS, and
many more.Input of \BWS\ is a set of initial states $I \subseteq
  \Sinf$, and a non-initial final state~$q$. The algorithm maintains a work
  set $W \subseteq \Sinf$ of unprocessed states, and a set $U \subseteq
  \Sinf$ of minimal encountered states. Starting from a final state $q$,
it successively computes minimal \emph{cover preimages}
\begin{equation}
  \label{equation: cover predecessors}
%  \format\hspace*{-9pt}\CovPre(w) =
  \min\{p : \exists w' \covers w : p \transrelinf w'\}
\end{equation}
where $\transrelinf$ is the transition relation of the WQOS. The search
terminates either by backward-reaching an initial state (thus proving
coverability of~$q$), or when no unprocessed state remains (thus proving
uncoverability).

Cover preimage computation is somewhat non-standard due to the use of the
covering relation \equationref[]{cover predecessors}. Moreover, performing
this step directly on a Boolean program, rather than on a transition
system, requires a backward ``execution'' of the program, an operation that
can only be implemented reasonably using the program control flow graph. On
the other hand, due to the sound-and-complete nature of this algorithm for
the broad class of WQOS, flavors of it are widely used in unbounded-thread
program verification (e.g.~\cite{KKW14a}). For these reasons we have added
dedicated support for this algorithm to \ourapi, in the form of an
implementation of the cover preimage operation \equationref[]{cover
  predecessors} applied directly to a multi-threaded Boolean program state.
The idea for this implementation of \equationref[]{cover predecessors} is
borrowed from \cite{LW14}.
}%end of appendix
\end{document}